\documentclass[manuscript, screen]{acmart}
\AtBeginDocument{%
  }

\setcopyright{acmlicensed}
\copyrightyear{2026}
\acmYear{2026}
\acmDOI{XXXXXXX.XXXXXXX}
\acmConference[Co-Data Workshop @ CHI2026]{2026 Co-Data: Cultivating Effective Human–LLM Collaboration for Collaborative Data Processing Workshop}{April 13, 2026}{Barcelona, Spain}
\acmISBN{978-1-4503-XXXX-X/2018/06}

\begin{document}

\title{Beyond One User and One Model: Rethinking LLMs as Mediators in Collaborative Sensemaking and Ideation}

\author{Pengqi Wang}
\affiliation{%
  \institution{York University}
  \city{Toronto}
  \state{Ontario}
  \country{Canada}}
\email{ericwang@yorku.ca}

\author{Emily Kuang}
\affiliation{%
  \institution{York University}
  \city{Toronto}
  \state{Ontario}
  \country{Canada}}
\email{ekuang@yorku.ca}

\renewcommand{\shortauthors}{Wang and Kuang}

\begin{abstract}
    As LLMs enter qualitative data workflows, they are predominantly designed for individual researchers rather than collaborative teams---risking disruption to the collective negotiation essential to sensemaking, driving premature consensus, and potentially erasing marginalized voices.
    In this position paper, we argue that LLMs must be reconceived as active team mediators. 
    Grounded in empirical studies of human-LLM collaborative analysis, we propose three design directions: quote-level traceability to preserve accountability of human and LLM contributions, controllable delegation boundaries to maintain researcher agency across workflow stages, and structured dissent mechanisms to surface interpretive conflicts that teams might otherwise suppress. 
    We ground these directions in accessibility research as a domain where methodological rigor and representational equity are obligatory.
\end{abstract}
    


\maketitle

\section{Introduction \& Motivation}

Complex knowledge work is fundamentally a social process. 
When research teams make sense of interview transcripts or brainstorm design concepts, they rely on negotiation, debate, and the reconciliation of different perspectives. But as Large Language Models (LLMs) are increasingly integrated into these qualitative data workflows, they are largely deployed as tools for individual analysts. 
This individualistic framing disrupts two distinct but deeply intertwined phases of collaborative research: the interpretive work of \textbf{sensemaking} and the generative work of \textbf{ideation}.

\textbf{Sensemaking} is the process of working through large volumes of qualitative data to identify patterns, construct themes, and build an interpretive account. In team settings, sensemaking is distributed: multiple analysts read overlapping subsets of the data, propose codes, negotiate disagreements about what a passage means, and iteratively refine a shared codebook. The quality of the final analysis depends on this negotiation.
Aside from data volume, the main challenge of collaborative sensemaking is the interpretive labor of reconciling divergent readings~\cite{feustonPuttingTools2021, kabbaraAIassistedSensemaking2025}.

\textbf{Ideation} moves in the opposite direction: from constraints toward possibilities, as teams explore alternatives, combines early findings, and evaluates emerging concepts.
Ideation often follows or interleaves with sensemaking: the patterns a team surfaces in the data become the raw material for creative exploration~\cite{chenCoExploreDSFraming2025, shenIdeationWebTracking2025}. However, the cognitive demands are distinct. Sensemaking rewards careful attention to the data; ideation rewards the willingness to move beyond it toward what could be~\cite{khanAutomationHow2025, yeDesignSpace2025}.

LLMs introduce distinct risks to both activities.
In \textbf{sensemaking}, the primary concern is premature closure: confident, LLM-generated codes can anchor interpretations before analysts engage deeply with the raw data, shifting their cognitive load from generative negotiation to passive evaluation~\cite{kabbaraAIassistedSensemaking2025}. 
In \textbf{ideation}, the risk is premature convergence: early, polished concepts can cause teams to rally around a narrow design space instead of exploring alternatives~\cite{suhLuminateStructured2024, khanAutomationHow2025}.
A growing body of HCI research addresses these concerns through individual-use tools that emphasize transparency for validation~\cite{gebreegziabherPaTATHumanAI2023, yeScholarMateMixedInitiative2025, overneySenseMateAccessible2024} or enforce structured exploration to maintain creative breadth~\cite{suhLuminateStructured2024, chenCoExploreDSFraming2025, shenIdeationWebTracking2025}.

Yet, both streams of work share a core limitation: 
modeling human-LLM interaction strictly as a one-on-one exchange ignores the messy reality of teamwork. 
This is exactly the gap that this workshop addresses: how LLMs can shape team dynamics around trust, control, and delegation in data-intensive collaborative work.
This gap matters most when the analyzed population is absent.
In accessibility research, for instance, teams analyze observations with people with disabilities, caregivers, and practitioners, turning lived experiences into design concepts~\cite{dasThatComes2024, bennettWhatPoint2020, dangolAIDrivenSupport2024, wuSpeechAI2025, almufarehConceptualModel2024}. If an LLM flattens these diverse voices into a generic consensus before any team debates, it actively erases those unique experiences.

In this position paper, we argue that LLMs in collaborative data workflows should be reconceived as mediators rather than individual tools. We present three design directions that follow this reframing, ground them in sensemaking and ideation theory, and close with open questions for the workshop community.

\section{Position: LLMs as Mediators, Not Just Tools}

Our core position is that LLMs in collaborative qualitative analysis should function as \textbf{mediators} within the team's process. Treating the model as a mediator rather than an individual tool carries three implications for how we design collaborative systems:

\textit{Mediation makes friction productive.} 
This friction is exactly what qualitative teams use to test their assumptions and build stronger insights~\cite{feustonPuttingTools2021}. In ideation, the value of multiple perspectives prevents premature convergence on any one direction~\cite{chenCoExploreDSFraming2025}. An LLM positioned as a mediator can actively surface these tensions. For instance, it could highlight where two researchers interpreted the same passage differently, or warn a team when they become fixated on a single idea during brainstorming. This is fundamentally different from a tool that each analyst consults independently, because the mediator's contributions are visible to and contestable by the entire team.

\textit{Mediation turns LLM limitations into a shared responsibility.} Every LLM-generated summary, theme suggestion, or design concept is shaped by the prompt, training distribution, and limited context windows. 
In a single-user tool, managing these biases and omissions is an individual's responsibility. 
In a team, it becomes a collective risk.
If one researcher accepts an LLM-generated theme without scrutiny, that hidden bias gets baked into the team's shared knowledge base.
A mediator framing requires that the LLM's contributions be inspectable by all team members, providing clear links to the evidence it used and explicitly flagging the data it excluded~\cite{yeDesignSpace2025, yeScholarMateMixedInitiative2025}.

\textit{Mediation acts as a structural safeguard for equity.} This brings us to our grounding application. In accessibility research, the participants whose data are being analyzed often belong to underrepresented communities~\cite{bennettWhatPoint2020, dasThatComes2024}. Their voices live in the transcripts and researchers' interpretations. An LLM that merges distinct disability experiences into a single theme, or ignores rare but crucial accounts, may silence those voices.
A mediator can mitigate the risk by keeping individual voices traceable through layers of abstraction. It can actively alert the team when a proposed theme obscures distinct minority experiences that the group might otherwise overlook~\cite{feustonPuttingTools2021}.

Our position is grounded in ongoing co-creation with accessibility researchers and community partners, including disability rights organizations and care practitioners. We will engage these partners in future studies to ensure that LLM constraints reflect the priorities of the communities whose data are at stake. This accessibility context will expose the coordination, transparency, and equity challenges any team-based human-LLM workflow will inevitably face.

\section{Theoretical Framing \& Design Directions}

Two theoretical traditions inform our proposed design directions. \textbf{Sensemaking} is an iterative process of structuring information where analysts must actively construct meaning~\cite{weickSensemakingOrganizations1995, braunUsingThematic2006, heRevisitingBa}. Because analysts resist early AI structure but welcome later validation~\cite{feustonPuttingTools2021}, LLM contributions must arrive as callable, provisional proposals~\cite{jarrahiGenerativeAI2025}. Conversely, \textbf{ideation} diverges as problem and solution spaces co-evolve~\cite{dorstCreativityDesign2001, macleanQuestionsOptions1996, schonReflectivePractitioner2017}, requiring structured exploration and traceable records to support divergent thinking rather than automating convergent decisions~\cite{chenCoExploreDSFraming2025, yeDesignSpace2025, khanAutomationHow2025}. Despite their different rhythms, both are vulnerable to confident, highly polished LLM outputs that discourage the slow, deliberate team engagement necessary for rigorous research. This shared risk motivates three design directions.

\subsection{Design Direction 1: Quote-Level Traceability}

Every LLM-generated theme, code, or summary should link back to the specific participant quotes that support it. This requirement is a precondition for meaningful team deliberation. For instance, if an LLM proposes merging different disability experiences under a single label, the team needs to inspect exactly which quotes the model relied on and decide whether the merge is justified. 
Furthermore, numerical grounding can complement qualitative traceability by providing quantitative data on how often an LLM-generated code was applied across the corpus~\cite{kabbaraAIassistedSensemaking2025}. In a team setting, quote-level traceability acts as a critical coordination mechanism. It ensures that all analysts can inspect the same evidence base rather than relying on verbal summaries of what the LLM suggested to each of them separately.

\subsection{Design Direction 2: Controllable Delegation Boundaries}

Researchers should be able to specify where the LLM can and cannot intervene, and these boundaries should be adjustable by workflow stage. %
In team settings, these boundaries also act as a pedagogical and governance tool. For example, a senior analyst might restrict LLM involvement during initial coding to guarantee junior members genuinely engage with the raw data, later relaxing those constraints when the team needs help managing a massive codebook~\cite{yeDesignSpace2025}.

\subsection{Design Direction 3: Structured Dissent Mechanisms}

The system must stop optimizing for single, coherent interpretations. It should actively surface counter-themes, outlier data points, and alternative framings that challenge the emerging consensus. When an AI produces a clean, well-organized output, teams tend to agree quickly, which can mask vital distinctions in the data~\cite{feustonPuttingTools2021}. Presenting multiple LLM-generated responses simultaneously~\cite{suhLuminateStructured2024} or a ``surprise me'' mode~\cite{kochImageSenseIntelligent2020} can prevent this fixation and broaden the exploration. Structured dissent is especially critical when analyzing disability data, where experiences are highly heterogeneous. 
Crucially, challenging a colleague's interpretation carries steep interpersonal risk, especially in hierarchical teams~\cite{dasThatComes2024}. If the system itself raises the disagreement as a routine part of the workflow, no individual team member has to bear the social cost of being the dissenter.

\section{Preliminary Findings \& Design Rationale}

Our proposed design directions build upon a series of empirical research examining human-AI collaborative analysis in usability evaluation~\cite{kuangCraftingHumanAI, kuangMergingResults2022, kuangCollaborationConversational2023, kuangEnhancingUX2024, kuangItBecame2026,kuangEvaluatingUsability2025, shenMultiUXHumanAI2026, wangExploringImpact2024}. Usability evaluation serves as a highly structured example of collaborative qualitative data work: evaluators observe sessions, interpret friction points, and negotiate merged findings. 
While these foundational studies examined broader AI systems, the coordination and trust dynamics they uncovered apply directly to LLMs. 
Across these studies, several core dynamics emerged that directly motivate our shift from an ``LLM-as-tool'' to an ``LLM-as-mediator'' framework.

\textbf{Trust builds over time.} 
In a multi-session study tracking evaluators over repeated collaborations with conversational AI assistants, we found that trust constantly adapts based on experience~\cite{kuangItBecame2026}. 
Evaluators' analytical strategies shifted with their trust, moving from a cautious approach to a deeply integrated workflow as they learned the AI's strengths and limitations. This evolving reliance underscores why delegation boundaries must be dynamic (Design Direction 2), allowing teams to adjust LLM autonomy as their project phases and comfort levels shift.

\textbf{The timing of LLM intervention dictates its impact.} 
Controlled experiments revealed that proactive AI suggestions triggered confirmation bias, whereas suggestions provided \textit{after} human analysis offered productive validation and increased trust~\cite{kuangEnhancingUX2024}. Because premature LLM intervention undermines both sensemaking and ideation, Design Direction 2 insists that delegation boundaries remain callable and dependent on phases.
This complementary breadth motivates Design Direction 3: the system should actively surface what the team is missing, functioning as a structured dissenting voice rather than confirming the team's existing frame.

\textbf{LLM restructures team coordination and social hierarchies.} 
Our international survey study highlighted that the primary bottleneck in team analysis is coordination \cite{kuangMergingResults2022}. 
Introducing LLM into this mix changes how team members interact.
LLM outputs can resolve misunderstandings, but they also create tension when the LLM's interpretation challenges human expertise, %
allowing whoever controls the LLM to dominate the discussion.
An LLM mediator must therefore make its evidence transparent through quote-level traceability (Design Direction 1) to ground team negotiations in shared data rather than assumed authority.

While these findings emerge from usability evaluations, they are not unique to that setting.
How trust shifts over repeated collaboration, how the timing of LLM input shapes what a team notices, and how LLM-generated content restructures informal hierarchies are concerns that any team engaged in collaborative qualitative research will eventually confront.
We are currently extending this line of work through formative studies with accessibility researchers and community partners. These studies involve semi-structured interviews and artifact walkthroughs of real analysis materials, including transcripts, codebooks, memos, and synthesis outputs, to document current collaborative workflows and produce concrete design requirements for LLM-mediated tools. %
The next phase will involve participatory co-design workshops to develop LLM-assisted collaborative tools, which will be evaluated through user studies.

\section{Open Questions for Discussion}

We close with questions arising from our empirical work and by a recent unresolved debate.

\textbf{Where is the line between structuring and interpreting?} Jowsey et al.~\cite{jowseyWeReject2025}, writing on behalf of over 400 qualitative researchers, recently rejected GenAI in all phases of reflexive qualitative analysis, arguing that meaning-making requires a subjective, positioned human researcher and that no statistical model can perform it. De Paoli~\cite{depaoliWhyWe2026} countered that this stance conflates replacement with augmentation: when a researcher remains in charge, GenAI functions as a partner that provokes better questions, not an interpretive authority that dictates answers.
Our proposed approach puts us right in the middle of this debate. We claim the LLM structures what the team sees without performing the interpretation itself. But if the AI decides which debates happen, isn't it already interpreting the data?
The Co-Data community is well-positioned to move this debate towards empirical design questions: how can we design interactions that protect human authority without quietly eroding it?

\textbf{How should interactions prevent LLM contributions from becoming default starting points?} Our previous studies showed that AI suggestions presented before human judgment triggered confirmation bias, while those presented afterward supported critical validation~\cite{kuangEnhancingUX2024}. 
In team settings, this compounds: if the LLM's proposed theme is presented identically to a senior researcher's interpretation, it may inherit unearned authority. What interactions can prevent teams from blindly trusting or instantly ignoring the AI?

\textbf{Who decides how much to rely on LLM when teammates trust it differently?} Our longitudinal evidence shows that trust in AI changes as users gain experience~\cite{kuangItBecame2026}. In a team, everyone adapts at their own pace. If the team sets the rules together, whose calibration prevails? If individually, how does the team stay aligned? This becomes especially messy in interdisciplinary teams where no single member holds methodological primacy.

\bibliographystyle{ACM-Reference-Format}
\bibliography{sample-base-bib}

\end{document}